\documentclass[usenatbib]{mn2e}
\usepackage{color,hyperref}
\definecolor{darkblue}{rgb}{0.0,0.0,0.3}
\hypersetup{colorlinks,breaklinks,
            linkcolor=darkblue,urlcolor=darkblue,
            anchorcolor=darkblue,citecolor=darkblue}
\usepackage{amsmath,amssymb}

\usepackage{txfonts}
\DeclareSymbolFont{cmletters}{OML}{cmm}{m}{it}
\DeclareMathSymbol{v}{\mathalpha}{cmletters}{"76}

\usepackage{graphicx}
\usepackage{amsmath} 

\defcitealias{mtb12b}{MTB12a}
\newcommand{\acknowledgements}{\section*{Acknowledgements}}

\newcommand{\be}{\begin{equation}}

\newcommand{\ee}{\end{equation}}

\newcommand{\abs}[1]{{\left|#1\right|}}
\newcommand{\dif}{\ensuremath{{\rm d}}}
\newcommand{\myvec}{\boldsymbol}
\newcommand{\Rlc}{\ensuremath{R_{\rm LC}}}
\newcommand\simless\lesssim   
\newcommand\simmore\gtrsim   

\newcommand\apj{\rmfamily{ApJ}}%
\newcommand\apjl{\rmfamily{ApJ}}%
\newcommand\aap{\rmfamily{A\&A}}%
\newcommand\mnras{\rmfamily{MNRAS}}%
\newcommand\aplett{\rmfamily{Astrophys.~Lett.}}%
\begin{document}
\label{firstpage}

\title[Time evolution of pulsar obliquity angle]{Time evolution of
  pulsar obliquity angle from 3D simulations of magnetospheres}
\author[A.~Philippov, A.~Tchekhovskoy, J.~G.~Li]{
Alexander Philippov$^1$\thanks{sashaph@princeton.edu},
Alexander Tchekhovskoy$^{2,3,4}$\thanks{Einstein Fellow},
Jason G.~Li$^1$
\\
$^1$Department of Astrophysical Sciences, Peyton Hall, Princeton University, Princeton, NJ 08544, USA\\
$^2$Lawrence Berkeley National Laboratory, 1 Cyclotron Rd,
  Berkeley, CA 94720, USA\\
$^3$Department of Astronomy, University of California Berkeley, Berkeley,
  CA 94720-3411, USA\\
$^4$Formerly at the Center for Theoretical Science, Jadwin Hall, Princeton University,
Princeton, NJ 08544, USA; Center for Theoretical Science Fellow}

\date{Accepted . Received ; in original form }
\pagerange{\pageref{firstpage}--\pageref{lastpage}} \pubyear{2013}
\maketitle

\begin{abstract}
The rotational period of isolated pulsars increases over time due to the
extraction of angular momentum by electromagnetic torques.
These
torques also change the obliquity angle
$\alpha$ between the magnetic and rotational axes. Although actual pulsar magnetospheres are plasma-filled, the
time evolution of $\alpha$ has mostly been studied for vacuum pulsar
magnetospheres.  In this work, we self-consistently
account for the plasma effects for the first time by analysing the results of
time-dependent 3D force-free and magnetohydrodynamic simulations
of pulsar magnetospheres.  We show that if
a neutron star is spherically symmetric and is embedded with a dipolar
magnetic moment, the pulsar evolves so as to minimise its spin-down
luminosity: both vacuum and plasma-filled pulsars
evolve toward the aligned configuration ($\alpha=0$). However, 
they approach the alignment in qualitatively different
ways. Vacuum pulsars come into alignment exponentially fast, with
$\alpha\propto \exp(-t/\tau)$ and $\tau\sim$ spindown timescale. In
contrast, we find that plasma-filled pulsars align much more slowly,
with $\alpha\propto
(t/\tau)^{-1/2}$.  We argue that the slow time evolution of obliquity
of plasma-filled pulsars
can potentially resolve several observational puzzles, including the
origin of normal pulsars with periods of $\sim1$ second, the evidence
that oblique pulsars come into alignment over a timescale of
$\sim10^7$ years, and the observed deficit, relative to an isotropic
obliquity distribution, of pulsars showing interpulse emission.
\end{abstract}


\section{Introduction}
\label{sec:introduction}

Rotation powered pulsars slow down as magnetised
winds carry away pulsars' rotational energy.
Although pulsar magnetospheres are filled
with plasma \citep{gol69} that produces observed radio emission
(see, e.g., \citealt{1998puas.book.....L,2008AIPC..983...29L,Radio} and references therein), most of the early work on understanding
the time evolution of pulsars was carried out
ignoring plasma effects, i.e., assuming a vacuum magnetosphere,  
primarily because a quantitative analytic solution was
readily available \citep{deutsch55,1970ApJ...160L..11G}. In this solution, in
addition to merely slowing down the star, magnetospheric torques also
lead to the alignment of pulsar magnetic and rotational axes.  In fact, this alignment process is so
efficient that vacuum pulsars become aligned before they have a chance
to substantially spin down \citep{Michel70}.  Presently, in the absence of a
self-consistent model, the evolution of pulsar obliquity with time
is usually neglected or considered in the framework of the vacuum
model that neglects plasma effects in the pulsar magnetosphere \citep{kaspi}.

Recent advances in axisymmetric \citep*{ckf99, gruzinov_pulsar_2005,
  mck06pulff, tim06} and oblique \citep*{spit06,kc09,petri12a,lst11,
  kalap12} force-free and magnetohydrodynamic (MHD)
\citep*{kom06,SashaMHD} as well as particle-in-cell (PIC,
\citealt{SashaPIC})
modelling of pulsar magnetospheres now allow
one to self-consistently account for plasma charges and currents that
substantially modify the magnetospheric structure and spin-down
torques. In this paper we show that the presence of plasma in the
magnetosphere also substantially affects the process of pulsar
alignment. We start with describing our numerical models in
\S\ref{sec:numerical-models} and derive pulsar evolutionary equations
in \S\ref{sec:torques}. We then calculate the torques and evolution of
pulsar obliquity angle in our vacuum magnetospheres in \S\ref{sec:vacuum-magn} and
plasma-filled magnetospheres in \S\ref{sec:ff-mhd-magn}. We derive the
relationship between spin-down and alignment torques in
\S\ref{sec:relat-spin-down} and conclude in \S\ref{sec:disc-concl}.

\section{Numerical Methods and Problem Setup}
\label{sec:numerical-models}

\begin{table}
\caption{Model details: model name, pulsar obliquity angle ($\alpha$),
  simulation resolution and the radius of the NS, $R_*$, in
  units of the light cylinder radius, $\Rlc$.}
\begin{tabular}{l|l|l|l}
\hline
Name & $\alpha[^{\circ}]$ & Resolution$^*$ & $R_*/\Rlc$  \\
\hline
\multicolumn{4}{c}{Relativistic MHD simulations (with HARM):}\\
HR01A0 & 0& $288\times 128 \times 1$&0.1 \\
HR01A45 & 45&$288\times 128 \times 256$ &0.1 \\
HR01A90 & 90& $288\times 128 \times 256$&0.1 \\
HR02A0 & 0& $256 \times 128 \times 1$&0.2 \\
HR02A15 & 15& $256 \times 128 \times 128$&0.2 \\
HR02A30 & 30& $256 \times 128 \times 128$&0.2 \\
HR02A45 & 45& $256 \times 128 \times 128$&0.2 \\
HR02A60 & 60& $256 \times 128 \times 128$&0.2 \\
HR02A75 & 75& $256 \times 128 \times 128$&0.2 \\
HR02A90 & 90& $256 \times 128 \times 128$&0.2 \\
HR0375A30 & 30& $256 \times 128 \times 128$&0.375 \\
HR0375A60 &60& $256 \times 128 \times 128$&0.375 \\
HR0375A90 &90& $256 \times 128 \times 128$&0.375 \\
HR05A30 & 30& $256 \times 128 \times 128$&0.5 \\
HR05A60 &60& $256 \times 128 \times 128$&0.5 \\
HR05A90 &90& $256 \times 128 \times 128$&0.5 \\
\multicolumn{4}{c}{Force-free models (with AS):$\dagger$}\\
ASFF0375A0 & 0& $1000\times1000\times1000$&0.375 \\
ASFF0375A15 & 15& $1000\times1000\times1000$&0.375 \\
ASFF0375A30 & 30& $1000\times1000\times1000$&0.375 \\
ASFF0375A45 & 45& $1000\times1000\times1000$&0.375 \\
ASFF0375A60 & 60& $1000\times1000\times1000$&0.375 \\
ASFF0375A75 & 75& $1000\times1000\times1000$&0.375 \\
ASFF0375A90 &90& $1000\times1000\times1000$&0.375 \\
\multicolumn{4}{c}{Vacuum models (with AS):$\dagger$}\\
ASVAC0375A0 & 0& $1000\times1000\times1000$&0.375 \\
ASVAC0375A15 & 15& $1000\times1000\times1000$&0.375 \\
ASVAC0375A30 & 30& $1000\times1000\times1000$&0.375 \\
ASVAC0375A45 & 45& $1000\times1000\times1000$&0.375 \\
ASVAC0375A60 & 60& $1000\times1000\times1000$&0.375 \\
ASVAC0375A60ext & 60& $1000\times1000\times1000$&0.375 \\
ASVAC0375A75 & 75& $1000\times1000\times1000$&0.375 \\
ASVAC0375A90 &90& $1000\times1000\times1000$&0.375 \\
\hline
\end{tabular}\\
$^\ast$  Given as $N_r \times N_{\theta} \times
  N_{\phi}$ for MHD and $N_x \times N_y \times N_z$ for vacuum and
  force-free models.\\
$\dagger$ In all vacuum and force-free models the stellar radius is resolved with $30$
cells, with the exception of ASVAC0375A60ext, in which it is resolved
with $15$ cells.
\label{models}
\end{table}

We carried out a number of time-dependent 3D simulations of oblique
pulsar magnetospheres in the relativistic MHD,
force-free and vacuum approximations.  The models start with a
perfectly conducting star of radius $R_*$ that rotates at an angular
frequency $\Omega=2\pi/P$, where $P$ is the pulsar period. The star is
embedded with a magnetic dipole field of dipole moment
$\myvec\mu$ that makes an angle $\alpha$ with the rotational
axis, $\myvec\Omega$.  The models are listed in
Table~\ref{models} and cover the full range of obliquity angles,
$\alpha=0^\circ{-}90^\circ$, and a range of angular frequencies,
$\Omega R_*/c \equiv R_*/\Rlc=0.1{-}0.5$, where $\Rlc =
c/\Omega$ is the light cylinder (LC) radius. We note that the rotation of
the star causes substantial deviations away from the initial dipolar
magnetic field, especially around and beyond the LC where special
relativistic effects become important \citep{bs09b,bs09a}.

We carry out our relativistic MHD simulations (Tab.~\ref{models}) using the
\texttt{HARM} code \citep{gam03,tch07,mb09,tch11,mtb12} in a spherical polar
grid, $r$, $\theta$, $\varphi$, but we sometimes also use the cylindrical
radius, $R = r\sin\theta$.
We denote these models as HRmmAnn, where mm represents the value of $R_*/\Rlc$ and nn
the value of $\alpha$ in degrees. 
We carry out our force-free and vacuum simulations (Tab.~\ref{models})
with the code by
\citet{spit06} in a Cartesian grid and denote the
force-free models as ASFFnnAmm and vacuum models as ASVACnnAmm. More
details of the 
setup and general properties of our models are described in
\citet{SashaMHD} for relativistic MHD simulations and in
\citet{spit06,lst11} for force-free and vacuum models.

\section{Torques and Their Effect on Pulsar Obliquity}
\label{sec:torques}

\begin{figure}
\begin{center}
    \includegraphics[width=0.7\columnwidth]{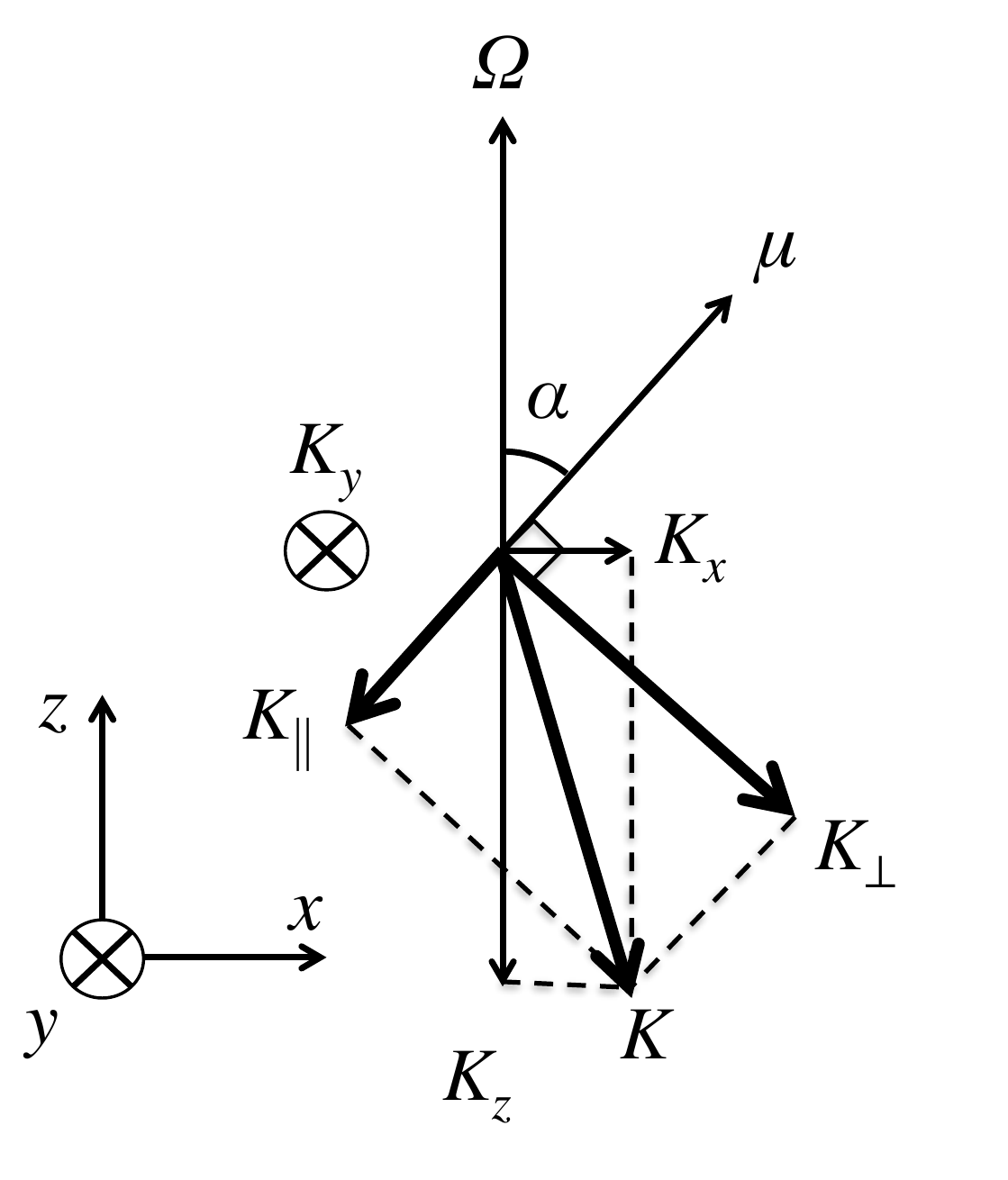}
  \end{center}
  \caption{Coordinate system used in this paper. Vector
    $\myvec{\Omega}$ shows the pulsar rotational axis, $\myvec{\mu}$
    the magnetic axis, and $\myvec K$ the magnetospheric torque on the
    NS. The $z$-component of the torque, $K_z$,
acts opposite to $\myvec\Omega$,
therefore it decreases the magnitude of $\myvec\Omega$ without
changing its direction, whereas $K_x$ and $K_y$ components act perpendicular to
$\myvec\Omega$. Therefore, while they keep the magnitude of
$\myvec\Omega$ the same, they change its orientation relative to
$\myvec\mu$. See \S\ref{sec:torques} for more detail. 
Description of $K_{\parallel}$ and $K_{\perp}$
is given in Section 6.}
\label{fig:axes}
\end{figure}

To compute the torques acting on the neutron star (NS), we choose an inertial
coordinate system illustrated in Figure
\ref{fig:axes}: $(\myvec
e_x, \myvec e_y, \myvec e_z)$, with the magnetic moment
$\myvec{\mu}$ instantaneously lying in the $x$-$z$ plane and $\myvec\Omega$
instantaneously pointing along the $z$-axis: $\myvec e_x =
[[\myvec e_{\Omega} \times \myvec e_{\mu}] \times
\myvec e_{\Omega}]$, $\myvec e_y = [\myvec e_{\Omega}
\times \myvec e_{\mu}]$, and $\myvec e_z = \myvec
e_{\Omega}$.  
The $i$-th component ($i=x,y,z$) of the torque acting on the star is given by \citep{LL}
\be
\label{eq:Ki}
K_i = -\int \varepsilon_{ijk} r_j T_{kl} {\rm d}A_l, 
\ee
where the integral is carried over the entire surface area of the
NS, summation over repeated indices is implied, 
$\varepsilon_{ijk}$ is the antisymmetric tensor, $\myvec
r=(x,y,z)$,  ${\rm d}A_l$
is the element of the surface area normal to $r_l$, and $T_{kl}$ are
the physical components of energy-momentum tensor. The minus sign
in eq.~\eqref{eq:Ki} appears because the angular momentum carried by the
outgoing wind is of the opposite sign to the rate of change of
the angular momentum of the NS.
In spherical coordinates, in which we
carried out our relativistic MHD simulations, 
eq.~\eqref{eq:Ki} becomes \citep{Michel70}:
\begin{subequations}
\label{eq:Ksph}
\begin{align}
K_x &= \int r(T_{r\theta}\sin \varphi + T_{r\varphi}\cos \theta \cos
\varphi) {\rm d} A, \label{eq:Ksph1}\\
K_y &= -\int r(T_{r\theta}\cos \varphi - T_{r\varphi}\cos \theta \sin \varphi) {\rm d} A,\\
K_z &= -\int r(T_{r\varphi}\sin \theta) {\rm d} A. \label{eq:Ksph3}
\end{align}
\end{subequations}
The evolution equations for the angular velocity of the pulsar are:
\be I\frac{{\rm d}{\myvec{\Omega}}}{{\rm d}{t}} = {\myvec
  K}, \ee where $I$ is the stellar moment of inertia.\footnote{For
  simplicity, we assume spherical symmetry of the star, e.g., we
  neglect possible effects of stellar oblateness caused by its rapid
  rotation. Non-sphericity of the NS can lead to non-trivial effects,
  which were carefully studied by \citet{Melatos} within a vacuum
  framework. We defer the analysis of similar effects within an MHD
  framework to future work. \label{ftn:sphericity}}  The $z$-component of the torque, $K_z$,
acts opposite to $\myvec\Omega\equiv \Omega \myvec e_z$,
therefore it decreases the magnitude of $\myvec\Omega$ without
changing its direction:
\begin{equation}
I \frac{{\rm d}\Omega}{{\rm d} t} = K_z. \label{motion1}
\end{equation}
In contrast, $K_x$ and $K_y$ components act perpendicular to
$\myvec\Omega$ (see Fig.~\ref{fig:axes}). Therefore, while they keep the magnitude of
$\myvec\Omega$ the same, they change its orientation relative to
$\myvec\mu$. 

\begin{figure}
\begin{center}
    \includegraphics[width=1\columnwidth]{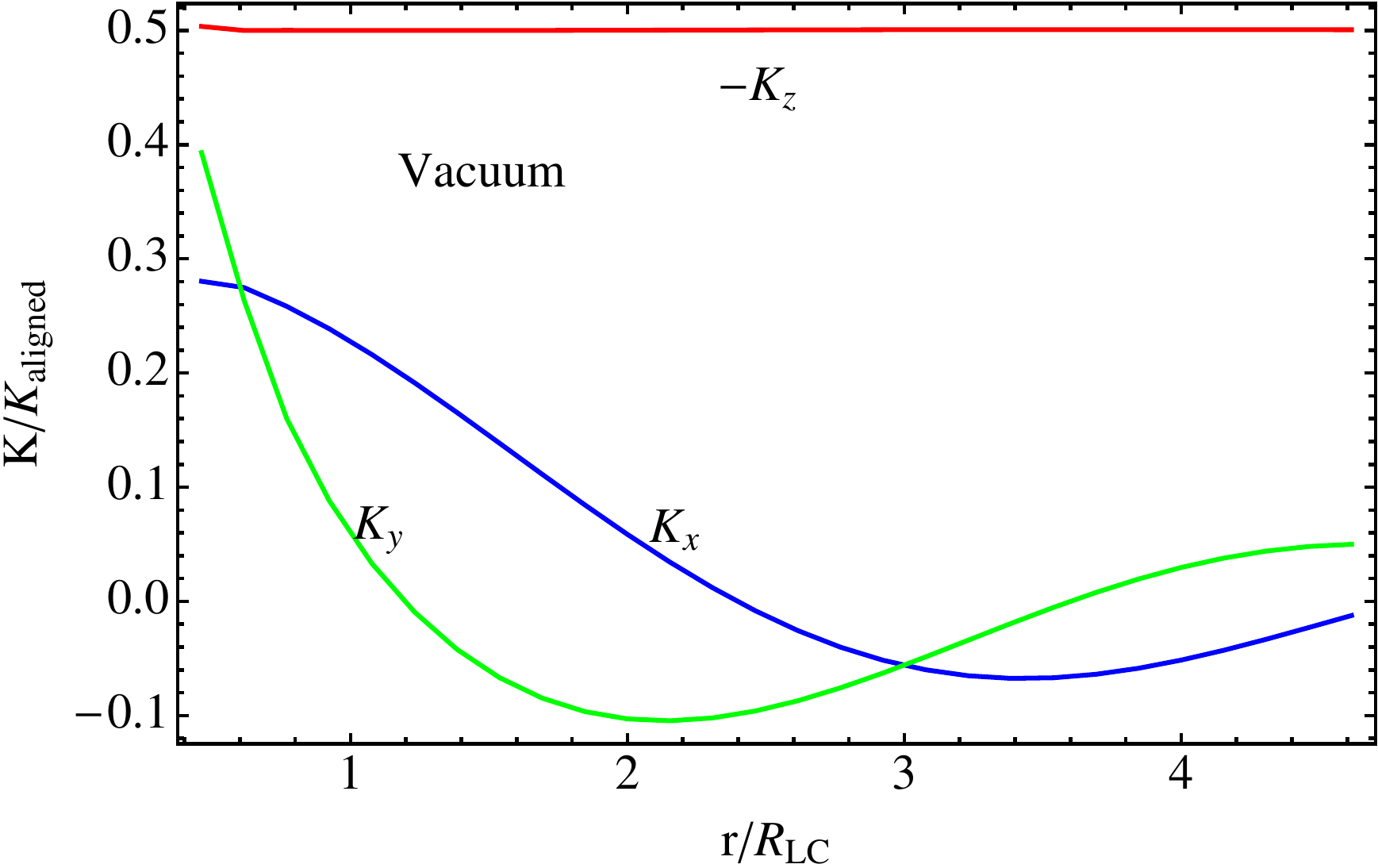}
  \end{center}
  \caption{Radial dependence of  torque components $K_x$, $K_y$ and
    $K_z$ calculated using the
    components of energy-momentum tensor in simulation ASVAC0375A60ext. The $z$-component of the magnetospheric
torque, $K_z$, is conserved and independent of the distance from the
star, so it can be measured at any distance. The two other
components of the torque, $K_x$ and $K_y$, show significant variations with
radius and should be measured at the surface of
the star, where the torque is transferred to the NS crust.}
\label{fig:Lxvac}
\end{figure}

\begin{figure}
\begin{center}
    \includegraphics[width=1\columnwidth]{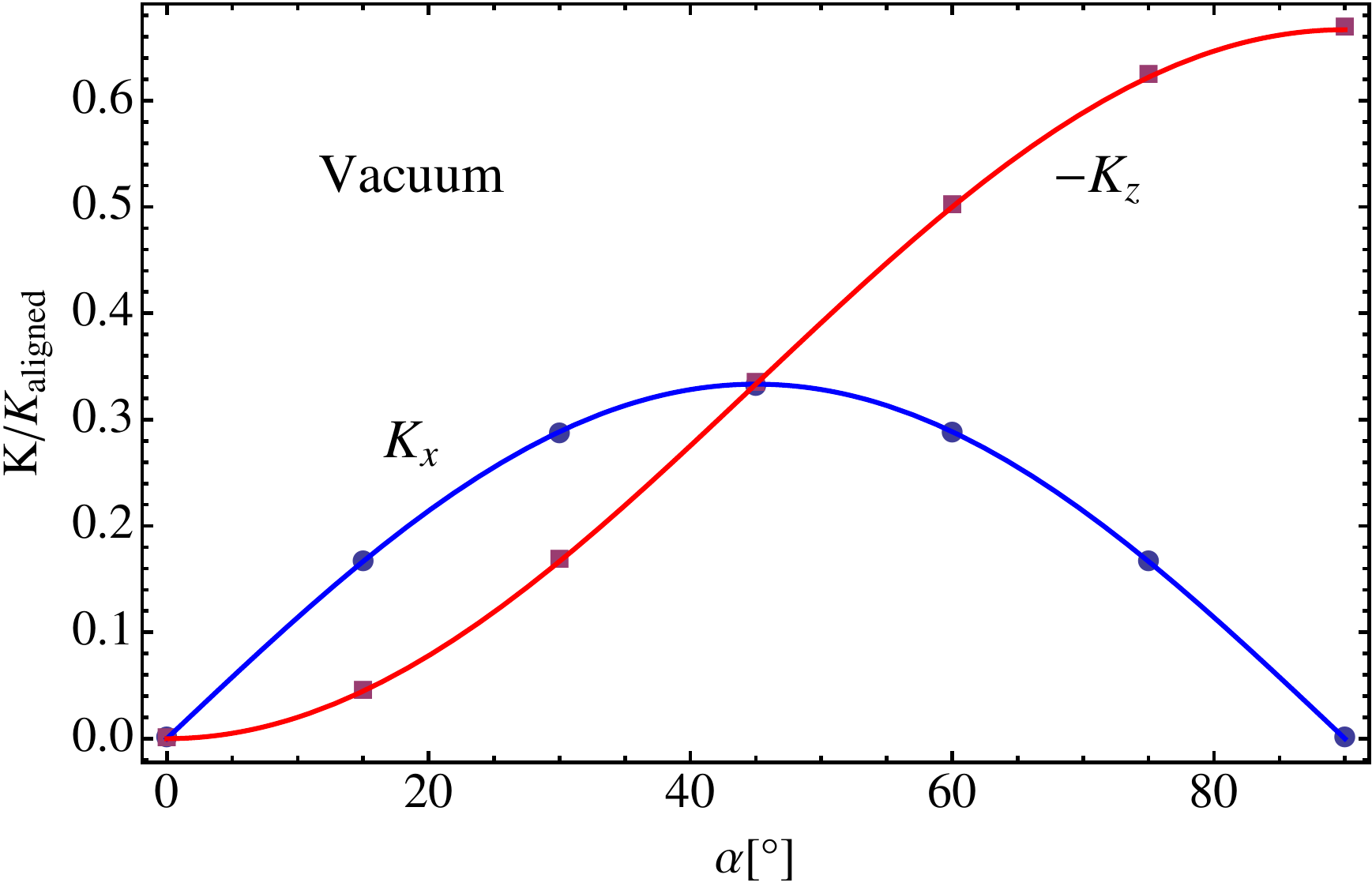}
  \end{center}
  \caption{The dependence of spindown torque $K_z$ (red squares)
    and alignment torque $K_x$ (blue circles)
    on pulsar obliquity angle $\alpha$ in vacuum models (see Tab.~\ref{models}). Solid lines show
    the analytic expressions (\ref{vacuum1})-(\ref{vacuum2}), whereas
    dots and squares show our numerical results 
    (models ASVAC0375A0--ASVAC037590). The agreement of the numerical results
    with the analytical expressions is excellent.}
\label{fig:vacuum}
\end{figure}

In particular, $K_x$, causes $\myvec\Omega$ to
rotate clock-wise in the $\myvec\mu{-}\myvec\Omega$ plane
(see Fig.~\ref{fig:axes}),
\begin{equation}
I\Omega \frac{{\rm d}\alpha}{{\rm d} t} = -K_x, \label{motion2}
\end{equation}
where we used $\dif\Omega_x/\dif t = -\Omega\dif\alpha/\dif t$ for
clock-wise rotation of $\myvec\Omega$ in $\myvec\mu{-}\myvec\Omega$
plane.  While it is tempting conclude that this causes the direction
of rotational axis to evolve, this is not so. This is because while
the vector $\myvec\Omega$ rotates in $\myvec\mu{-}\myvec\Omega$ plane
as described above, this plane itself rotates about $\myvec e_z$
together with the NS. If one averages over this rapid rotation of the
NS it is easy to see that the direction of $\myvec\Omega$ remains
almost unchanged as seen by a stationary observer,\footnote{In fact,
  the direction of $\myvec\Omega$ slightly oscillates. This causes
  higher-order corrections to the magnetospheric torques. For a
  perfectly spherical star studied in this paper these higher-order
  corrections can be neglected.} and it is the direction of
$\myvec\mu$ that is changing according to eq.~\eqref{motion2}.

Analogously, $K_y$-component causes $\myvec\Omega$ to rotate out
of $\myvec\mu{-}\myvec\Omega$ plane. As a result, the tip of
the vector $\myvec\Omega$ 
traces out a circle on the surface of the NS (this circle is
centred at the magnetic pole) and does so at an angular frequency
$\Omega_{\rm prec}$ given by
\begin{align}
I\Omega \Omega_{\rm prec} &= -\frac{K_y}{\sin \alpha},\label{motion3}
\end{align}
where we used $\dif\Omega_y/\dif t = -\Omega\Omega_{\rm prec}$ for
rotation of $\myvec\Omega$ out of the $\myvec\mu{-}\myvec\Omega$ plane.
As we neglect stellar
non-sphericity (see footnote \ref{ftn:sphericity}), eqs.~\eqref{motion1} and \eqref{motion2} are decoupled
from eq.~\eqref{motion3} and can be solved independently. We do so
in \S\ref{sec:vacuum-magn} for vacuum and in \S\ref{sec:ff-mhd-magn}
for force-free and MHD magnetospheres.

\section{Vacuum Pulsar Magnetospheres}
\label{sec:vacuum-magn}

Using an exact solution for the magnetospheric structure for a small
stellar radius in vacuum \citep{deutsch55}, it is straightforward to
compute the surface magnetospheric torque (eq.~\ref{eq:Ki}) on the NS
\citep{Michel70}:
\begin{align}
K_x &= \phantom{-}\frac{2}{3}K_{\rm aligned}\sin \alpha \cos \alpha, \label{vacuum1}\\
K_z &= -\frac{2}{3}K_{\rm aligned}\sin^2 \alpha, \label{vacuum2}
\end{align}
where, as we discuss in \S\ref{sec:ff-mhd-magn}, 
\be
K_{\rm aligned} = \frac{\mu^2 \Omega^3}{c^3}
\label{eq:kaligned}
\ee
is the spin-down torque
of an aligned force-free or MHD pulsar. 

\begin{figure*}
\begin{center}
    \includegraphics[width=2.\columnwidth]{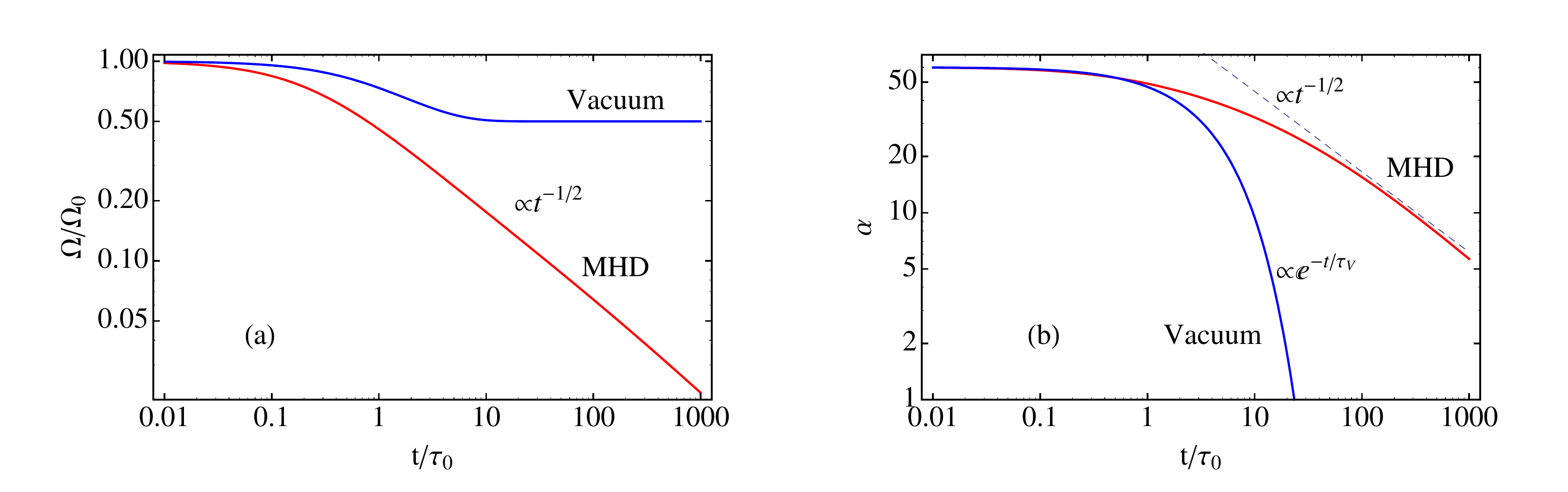}
  \end{center}
  \caption{{\bf [Left panel]:} Time evolution of
    pulsar angular frequency, $\Omega$,
    for vacuum (blue curve) and
    MHD (red curve) 
    pulsars for an initial inclination angle $\alpha_0= 60^{\circ}$. 
    When specialised to a
    fiducial initial period $P_0 = 10\ \rm{ms}$ and magnetic field
    strength of $B_0 = 10^{12}$~G,
    the horizontal axis is scaled to $\tau_0 = 10^4$~yr, see
    eq.~\eqref{eq:tau0}.
    Whereas vacuum pulsars settle to a constant rotation rate
    $\Omega_\infty=\Omega_0
    \cos \alpha_0=\Omega_0/2$ on a timescale that is of order of
    pulsar spindown time $\tau_0$,
    plasma-filled (MHD) pulsars continue to spin down as a power-law
    in time,
    $\Omega\propto t^{-1/2}$. The
  continued decrease of $\Omega$ in plasma-filled magnetospheres can
  help one to explain the origin of normal pulsars with periods of order
  of one second (see \S\ref{sec:ff-mhd-magn} for a discussion). {\bf [Right
    panel]:} The time evolution of pulsar obliquity angle for vacuum
  (blue) and MHD (red) pulsars. Vacuum
  pulsars evolve exponentially fast to an aligned configuration, 
  while for MHD pulsars the obliquity angle decreases to
  zero as a power-law, $\alpha \propto t^{-1/2}$. The dashed line shows the
  asymptotic behaviour given by equation (\ref{eq:alphamhd}).  For
  the fiducial value, $P_0 = 10$~ms, the alignment time for an
    MHD magnetosphere is $\simeq100\tau_0\approx10^6$ yr.}  
\label{fig:evolution}
\end{figure*}

As a comparison to the analytic model, we carried out numerical vacuum
magnetosphere simulations for a wide range of obliquity values (models
ASVACxxx, see Tab.~\ref{models}).  For instance, Fig.~\ref{fig:Lxvac}
shows the radial run of the components of $\myvec K$, which we
computed according to eq.~\eqref{eq:Ki}, for our fiducial vacuum model
VAC0375A60ext, which is a simulation with an obliquity angle $\alpha =
60^\circ$ and the period such that $R_*/\Rlc = 0.375$.\footnote{In
  fact, the torques do depend on $R_*/\Rlc$, as discussed by
  \citet{Melatos97}. For our computational star these corrections are
  of order of $(R_*/\Rlc)^2$ and affect the value of the torques by
  $\sim 5$ per cent. This is of order of the accuracy of our
  numerical solution, so, for simplicity, we neglect these
  corrections.}  As expected, the $z$-component of the magnetospheric
torque, $K_z$, is conserved and independent of the distance from the
star (this is a manifestation of the conservation of the $z$-component
of angular momentum). For this reason, the value of $K_z$ can be
measured at essentially any distance from the star: e.g.,
$K_z/K_{\rm aligned}\approx -0.5$ in this simulation, as can be seen in
Fig.~\ref{fig:Lxvac}.  The two other components of the torque, $K_x$
and $K_y$, show sinusoidal-like variations with radius,\footnote{These
  variations are an imprint of the periodic angular structure of an
  oblique magnetosphere that is carried out by the magnetospheric
  outflow.}  and it is important that we measure these components at
the correct location --- the stellar surface --- at which the
magnetospheric torques are transferred to the NS crust.

Since the radial profile of $K_x$ becomes flat at small $r/\Rlc$, we
can reliably measure its value at the NS surface, $K_x(R_*/\Rlc)\equiv
K_x(\Omega R_*/c)$. This also means that this
value is essentially independent of the pulsar period (which sets
$R_*/\Rlc$):  e.g., Fig.~\ref{fig:Lxvac} shows that
$K_x(r/\Rlc\ll1)/K_{\rm aligned}\approx 0.28$.
In contrast, $K_y$ approaches small radii at a
steep slope (see Fig.~\ref{fig:Lxvac}). This is because $K_y(R_*)$ diverges 
as $R_*/\Rlc\to0$ \citep{1970ApJ...159L..81D}. Due to this,
extra care is needed to measure $K_y$ at the stellar
surface in numerical models. In this paper we
concentrate on $K_x$ and $K_z$ torque components and do not
carry out any $K_y$ measurements.

Figure~\ref{fig:vacuum} shows that the analytic expressions for torque
components, $K_x$ and $K_z$, which are given by eqs.~\eqref{vacuum1} and
\eqref{vacuum2}, are in excellent agreement with our numerical vacuum
results, which are shown with discrete data points in
Fig.~\ref{fig:vacuum} (models ASVACxxAyy, see Tab.~\ref{models}).  To determine the time
evolution of pulsar's $\Omega$ and $\alpha$, we need to solve the system of
eqs.~\eqref{motion1} and \eqref{motion2} coupled with the vacuum
expressions for the torque, eqs.~\eqref{vacuum1} and \eqref{vacuum2}.

It is easy to show that the quantity,
\begin{equation}
\Omega\cos\alpha = \Omega_0 \cos\alpha_0,  \label{eq:cons_vacuum}
\end{equation}
is an
integral of motion, where the subscript ``$0$'' refers to the values
of pulsar variables at an initial time, $t=t_0$. 
One can also obtain the following analytic solution for the time
evolution of $\Omega$ and $\alpha$ \citep{Michel70}:
\begin{align}
\label{eq:omegavac}
\Omega &= \Omega_0\left\{1+\left[1-\exp\left(-{2 t}/{\tau^{\rm
          vac}_{\rm align}}\right)\right]\tan^2\alpha_0\right\}^{-1/2}\\
\label{eq:alphavac}
\sin\alpha &= \sin \alpha_0 \exp \left(-{t}/{\tau^{\rm
          vac}_{\rm align}}\right),
\end{align}
where $\tau^{\rm vac}_{\rm
  align}=1.5\tau_0\cos^{-2}\alpha_0$ is the alignment timescale of a
vacuum pulsar and 
\begin{equation}
\tau_0 = \frac{I c^3}{\mu^2 \Omega^2_0} \approx 10^4 \left(\frac{B_0}{10^{12}\, {\rm
    G}}\right)^{-2} \left(\frac{P_0}{10\, {\rm ms}}\right)^{2}~{\rm
yr}
\label{eq:tau0}
\end{equation}
is the characteristic spindown time of a pulsar.

From eq.~\eqref{eq:cons_vacuum} it is immediately clear that for
pulsar's rotation to appreciably slow down from its initial value, $\Omega\ll\Omega_0$,
one must start with a nearly orthogonal pulsar,
\begin{equation}
\label{eq:vac_limits}
\cos\alpha_0 = \frac{\Omega}{\Omega_0}\cos\alpha \le
\frac{\Omega}{\Omega_0} \ll 1.
\end{equation}
For instance, for a $10$-millisecond pulsar to end up as a normal $1$-second
pulsar, the pulsar must have been born with an extreme obliquity,
$\alpha_0\ge89.4^\circ$ (see also \citealt{Michel70}), which is
unlikely if
dipole moment orientation is random at birth.

The solid blue line in the left panel of Fig.~\ref{fig:evolution} illustrates that
the period of a pulsar
of a characteristic initial obliquity (we took $\alpha_0=60^\circ$ as
an example), can increase
in its lifetime by at most a factor of few: its
rotational frequency levels off at $\Omega = \Omega_0\cos\alpha_0=0.5\Omega_0$ and
obliquity vanishes, $\alpha=0$,  at $t\gtrsim
\tau^{\rm vac}_{\rm align}=6\tau_0$ (see
eqs.~\ref{eq:omegavac}--\ref{eq:alphavac}; the numerical evaluation
was performed for $\alpha_0 = 60^\circ$), in agreement with the blue
curve in the left panel of Fig.~\ref{fig:evolution}.
Therefore, if most pulsars start out as rapid rotators (with periods
of few to tens of milliseconds) with a random orientation of
their magnetic dipole moment, it is difficult to reconcile the
observed abundance of normal pulsars
(with periods of seconds) and the paucity of such pulsars
expected in the vacuum approximation.

\citet{1970ApJ...160L..11G} found that such a
rapid evolution of pulsar obliquity toward alignment does not occur if
stellar non-sphericity is taken into account, which makes it easier to
reconcile the expected pulsar period distribution with the observed one.
However, this and other studies \citep[e.g.,][]{Michel70,Melatos} on pulsar alignment rely on the vacuum
approximation to describe pulsar magnetospheres.
This is a rough description of magnetospheres in
nature because it neglects all plasma effects, in particular charges
and currents, that play a major role in shaping the magnetospheric
structure (e.g., \citealt{spit06}).  How does pulsar
obliquity evolution change once plasma effects are accounted for?

\section{Plasma-filled Pulsar Magnetospheres}
\label{sec:ff-mhd-magn}

In this section we compute magnetospheric torques and pulsar obliquity
evolution while accounting for the plasma effects on the pulsar
magnetosphere.\footnote{Force-free approximation makes the assumption
  that plasma inertia is negligible. This is an excellent
  approximation for NS magnetospheres \citep{gol69}, except perhaps
  the magnetospheric current sheet \citep{kom06,SashaMHD,SashaPIC}.  This
  approximation necessitates that the net force in the bulk of the
  plasma must vanish: if it did not, any non-compensated force acting
  in the bulk of the plasma would cause infinite acceleration.  This
  does not in any way prevent the magnetosphere from exerting torques
  on the NS via the Lorentz force that is due to surface currents
  and charges.}  In this work, we neglect the effects of the possible
non-sphericity of the NS (see footnote~\ref{ftn:sphericity}) and
higher order multipole components of the stellar field.

\begin{figure}
\begin{center}
    \includegraphics[width=1\columnwidth]{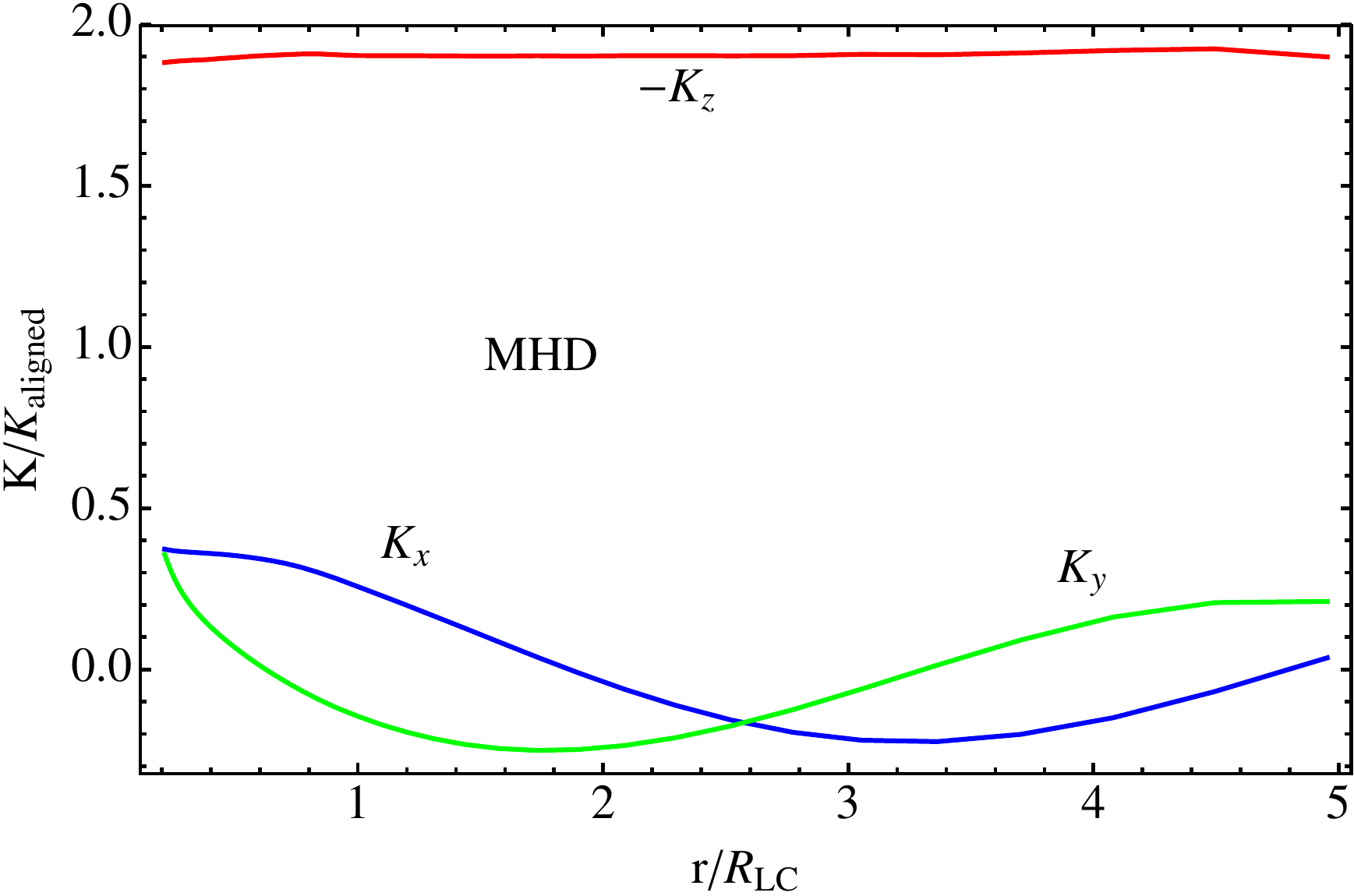}
  \end{center}
  \caption{The radial profile of torque components $K_x$, $K_y$ and $K_z$ 
    calculated using the components of
    energy-momentum tensor in our fiducial MHD simulation
    HR02A60. Similarly to the vacuum case (Fig.~\ref{fig:Lxvac}), 
    the $z$-component of the magnetospheric
torque, $K_z$, is conserved and independent of the distance from the
star, so it can be measured at any distance. The two other
components of the torque, $K_x$ and $K_y$, vary with
radius, and need to be measured at the surface of
the star, where the torques are transferred to the NS crust.}
\label{fig:LxMHD}
\end{figure}

We have carried out a number of relativistic MHD simulations for a full range of
pulsar obliquity, $\alpha=0{-}90^\circ$, and a rather wide range of pulsar period,
$R_*/\Rlc = 0.1{-}0.5$ (models HRxxAyy, see
Tab.~\ref{models}).  The results of our force-free models (models
ASFFxxAyy in Tab.~\ref{models}) are similar
to MHD and are not shown. Radial dependence for magnetospheric torque
components in our fiducial MHD model, HR02A60, which has $\alpha =
60^\circ$ and $R_*/\Rlc = 0.2$, is
shown in Fig.~\ref{fig:LxMHD}.  

The comparison to our fiducial vacuum
model, which is shown in Fig.~\ref{fig:Lxvac}, indicates that there are
many qualitative similarities between the two models:
$K_z$ is conserved as a function of distance, whereas $K_x$ and
$K_y$ show sinusoidal-like variations with distance. Quantitatively, however,
several clear differences emerge: we have $K_z/K_{\rm aligned}\approx-2$ in the MHD case as
opposed to $K_z/K_{\rm aligned}\approx-0.5$ in the vacuum case, 
a manifestation of a larger spin-down of plasma-filled (force-free and
MHD)  magnetospheres in comparison to vacuum magnetospheres 
\citep{spit06}. 
Due to this, the relative importance of the alignment torque, or the
ratio $K_x/\abs{K_z}$, is a factor of a few smaller in our fiducial MHD
simulation, $\approx0.2$, than in the vacuum one, $\approx0.6$. This
indicates that the presence of the plasma in the magnetosphere,
through the associated charges and currents, suppresses the alignment
torque relative to the spin-down torque. Therefore, we might expect that
pulsar obliquity angle evolution in plasma-filled magnetosphere case
(in MHD and force-free approximations) will be slower than in the
vacuum case.

\begin{figure}
\begin{center}
    \includegraphics[width=1\columnwidth]{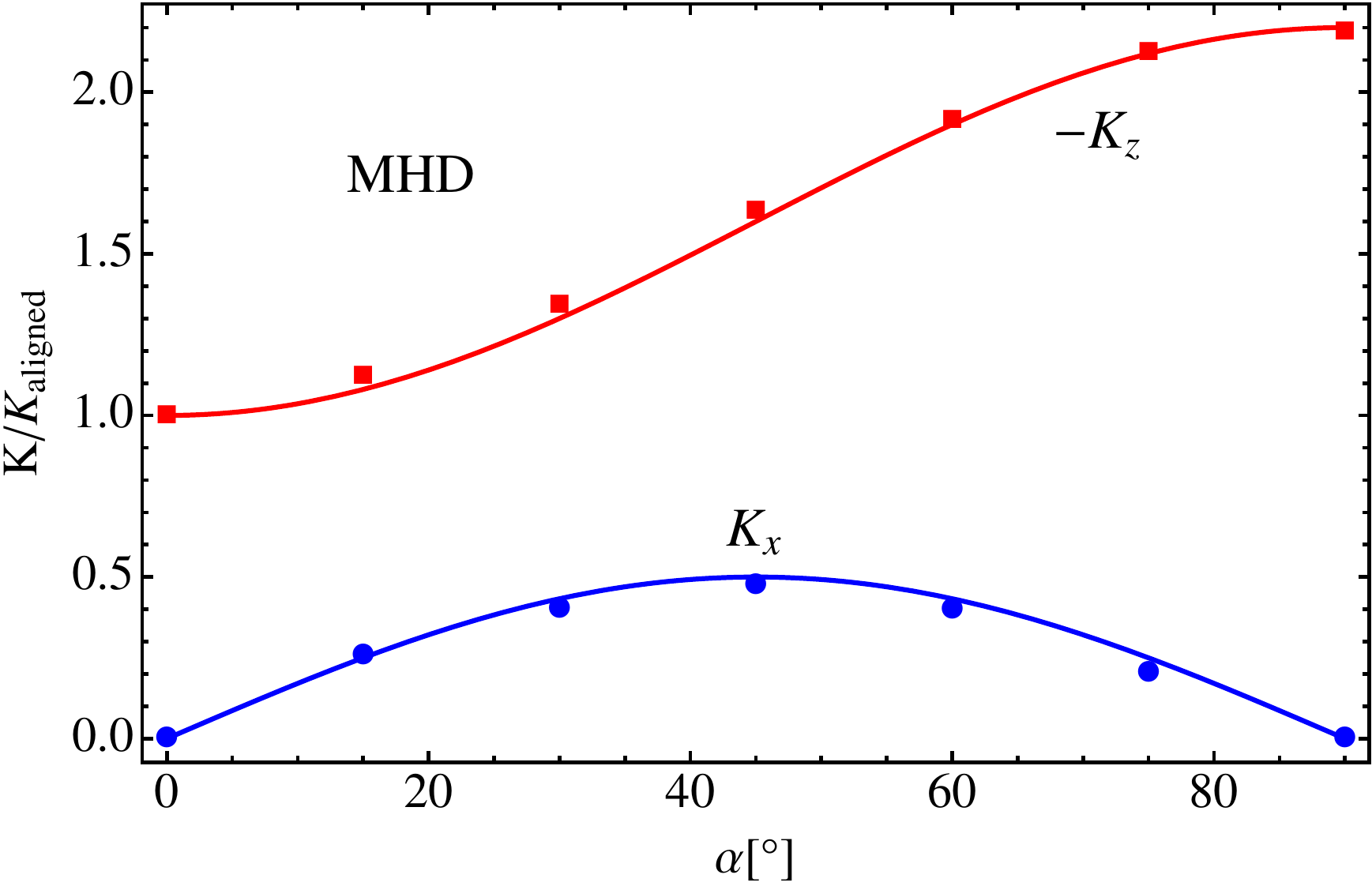}
  \end{center}
  \caption{The dependence of spindown torque $K_z$ (red curve)
    and alignment torque $K_x$ (blue curve)
    on pulsar obliquity angle $\alpha$  in our MHD models
    with $R_{*}/\Rlc = 0.2$. Solid lines show  fitted expressions
    (\ref{ff1})-(\ref{ff2}) with $k_1 \approx 1.2$ and $k_2 \approx
    1$, whereas dots and squares show our numerical results (models HR02A0--HR02A90).}
\label{fig:FF}
\end{figure}

\begin{figure}
\begin{center}
    \includegraphics[width=1\columnwidth]{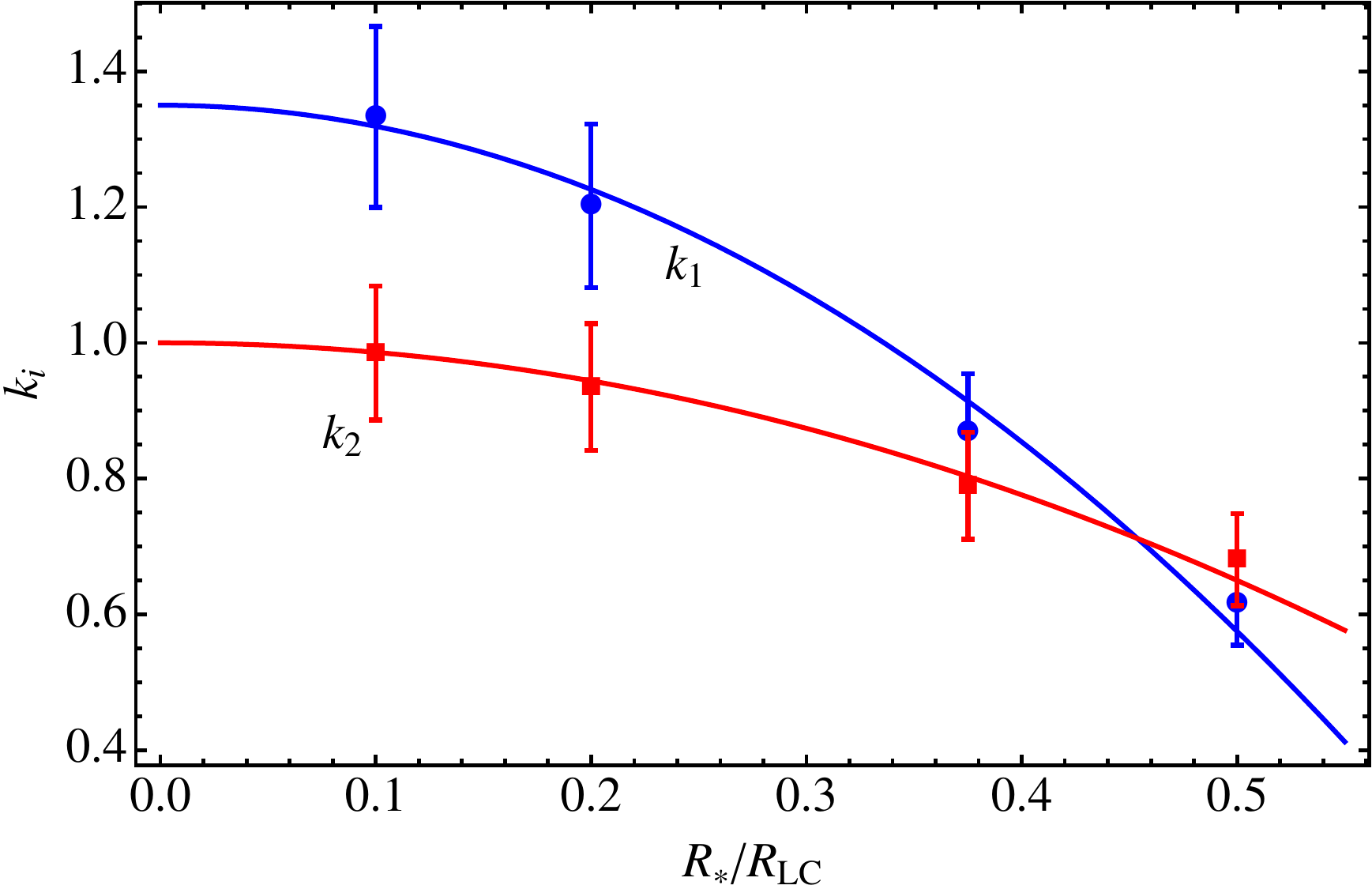}
  \end{center}
  \caption{The dependence on the radius of the pulsar of the coefficients
    $k_1$ and $k_2$ that enter eqs.~\eqref{ff1}--\eqref{ff2}.
    To determine their values most accurately, we 
    measured the value of $k_1$ from the simulations with $\alpha = 90^\circ$
    (blue circles) and the value of $k_2$ from the
    simulations with $\alpha = 45^{\circ}$
    (red rectangles). Red and blue lines are parabolic fits to the
    numerical data points. 
    The shown error bars at the level of $10$ per cent approximately indicate
    the uncertainties due to the measurements of the torque on
    the stellar surface and the differences in the values of $k_1$ and $k_2$
    obtained in the simulations carried out at different obliquity angles.}
\label{fig:k1}
\end{figure}

To test this hypothesis, we carried out MHD simulations for a full
range of obliquity angles. Figure~\ref{fig:FF} shows our MHD results
for a particular value of pulsar period, $R_*/\Rlc = 0.2$. They are
well-described by the following fits:
\begin{align}
K_x &= k_2 K_{\rm aligned}\sin \alpha \cos \alpha, \label{ff1}\\
K_z &= -K_{\rm aligned}\left(k_0 + k_1\sin^2 \alpha\right), \label{ff2}
\end{align}
where $k_0\approx k_2 \approx 1$ and $k_1\approx1.2$.
Interestingly, since $K_x>0$, MHD pulsars
evolve toward alignment, just as their vacuum counterparts. As we will
see below, however, they approach alignment along very different paths.
Note that the system of equations (\ref{ff1})--(\ref{ff2}) reduces to
that for the vacuum case (eqs.~\ref{vacuum1}--\ref{vacuum2}) for $k_0 = 0$, $k_1=k_2=2/3$.

Figure~\ref{fig:k1} shows that $k_i$, $i=0,1,2$, exhibit a weak
dependence on $R_*/\Rlc$: $k_0\approx1$ is virtually independent of
$R_*/\Rlc$ (not shown in Fig.~\ref{fig:k1}) and
$k_1$ and $k_2$ display a relatively weak dependence on $R_*/R_{\rm
  LC}$. Given the symmetry of the problem with respect to the
direction of rotation (the sign of $\Omega$), we expect that only even
powers of $R_*/\Rlc\equiv \Omega R_*/c$ enter into the expressions for
$k_i$.  However, given the limited range in $R_*/\Rlc$ accessible to
us numerically (the simulations become prohibitively expensive at
small values of $R_*/\Rlc$), we cannot robustly constrain the complete
functional form of $k_1$ and $k_2$ as a function of $R_*/\Rlc$. 
We see from Fig.~\ref{fig:k1} that the
following parabolic dependence describes the numerical results
reasonably well:
\begin{align}
  k_1 &\approx 1.4-3.1\times(R_*/\Rlc)^2, \label{eq:k1}\\
  k_2 &\approx 1.0-1.4\times(R_*/\Rlc)^2. \label{eq:k2}
\end{align}
The
origin of this dependence will be studied in future work. However, as
deviations from unity are not very large, we will sometimes for
convenience neglect this weak spin-dependence of $k_1$ and $k_2$.

We now combine the expressions for torques in the MHD approximation,
eqs.~\eqref{ff1} and \eqref{ff2}, with the evolutionary equations
\eqref{motion1} and \eqref{motion2}.  Under the assumption that factors $k_1$
and $k_2$ are constant, we find that the quantity,
\be 
\label{eq:cons_mhd}
\Omega \left(\frac{\cos^{k_1 + 1}\alpha}{\sin\alpha}\right)^{1/k_2} =
\Omega_0 \left(\frac{\cos^{k_1 +
      1}\alpha_0}{\sin\alpha_0}\right)^{1/k_2}, 
\ee 
is conserved in the
evolution of a pulsar in both the MHD and force-free approximations. 
We will see below that eq.~\eqref{eq:cons_mhd} imposes less rigidity
on pulsar time evolution as compared to the
vacuum counterpart, eq.~\eqref{eq:cons_vacuum}.

If we approximately take
$k_0 \approx k_1 \approx k_2 \approx 1$, we can derive an analytic
solution for the evolution of pulsar obliquity angle, $\alpha$:
\be
\label{eq:alphamhd}
\frac{1}{2\sin^2\alpha} + \log(\sin\alpha) = \frac{t}{\tau^{\rm MHD}_{\rm align}} + \frac{1}{2\sin^2\alpha_0}+\log(\sin\alpha_0),
\ee
where $\tau^{\rm MHD}_{\rm align} = \tau_0\sin^2 \alpha_0/\cos^4
\alpha_0$ is MHD pulsar alignment timescale. 

Figure \ref{fig:evolution} compares the evolution of $\Omega$ and $\alpha$
for an initial obliquity of $\alpha_0=60^{\circ}$ for MHD models with $k_1=1.4$
and $k_2=1$, as consistent with approximations given by
eqs.~\eqref{eq:k1}--\eqref{eq:k2} in the limit $R_*/R_{\rm LC}\to0$.  
From Fig.~\ref{fig:evolution}(b) it is clear that
asymptotically late in time, at $t \gg \tau^{\rm MHD}_{\rm
  align}=12\tau_0$ (the numerical evaluation is for
$\alpha=60^\circ$), MHD pulsar obliquity decreases as a power-law,
$\alpha \propto t^{-1/2}$ (see eq.~\ref{eq:alphamhd}).  As this is
much slower than in vacuum, where $\alpha\propto \exp(-t/\tau^{\rm
  vac}_{\rm align})$ (see eq.~\ref{eq:alphavac}), this confirms our
expectation that pulsar alignment in MHD magnetospheres occurs at a
much slower rate than in vacuum magnetospheres. 

As we discussed in \S\ref{sec:vacuum-magn}, a concerning aspect of
vacuum pulsars is that their period generally increases
only by at most a factor of a few and asymptotes to a constant value,
$P\simeq{\rm few}\times P_0$, which is generally too small to be
consistent with periods of a few seconds commonly seen in normal
pulsars. In contrast, in the MHD approximation pulsar period increases
as a power-law, $P\simeq P_0 (t/\tau_0)^{1/2}$, with virtually no
upper limit. This makes it much more comfortable to explain the
formation of normal pulsars as initially born as rapidly rotating
pulsars with periods of few to tens
of millisecond (see
\S\ref{sec:vacuum-magn}).

Why do plasma-filled and vacuum pulsars have such different
time-evolution? The key difference is in the spin-down rate of an
aligned rotator. For
instance, since a nearly aligned ($\alpha\ll1$) vacuum pulsar 
essentially does not spin
down, 
$\Omega\propto t^0$, due to eqs.~\eqref{motion2},
\eqref{eq:kaligned} and \eqref{ff1}, we get $d\alpha/dt \propto \alpha
\Omega^2 \propto \alpha t^0$, giving us the exponential decay of the
obliquity angle, $\alpha \propto \exp(-t/\tau)$. In contrast, since a
nearly aligned ($\alpha\ll1$) plasma-filled
pulsar spins down as $\Omega\propto t^{-1/2}$, we
get $d\alpha/dt \propto \alpha
\Omega^2 \propto \alpha t^{-1}$, thus giving us the power-law evolution, $\alpha
\propto (t/\tau)^{-\beta}$, where $\beta$ is an integration constant.  
To determine the value of $\beta$, we divide
eq.~\eqref{motion2} by eq.~\eqref{motion1} and with the help of \eqref{ff1} obtain:
\begin{equation}
  \label{motion12}
  \frac{{\rm d}\alpha}{\alpha} = k_2 \frac{{\rm d}\Omega}{\Omega}.
\end{equation}
Hence, $\alpha \propto \Omega^{k_2} \propto t^{-k_2/2}$.
  Since
$k_2\approx1$ for $R_*/\Rlc \ll 1$ (see eq.~\ref{eq:k2}), we can approximately write $\alpha \propto
t^{-1/2}$. 

Note the extreme sensitivity of obliquity angle to the value of the
alignment torque: $k_2$ enters into the $\alpha(t)$ dependence as a
power-law index. Thus, for a normal pulsar of period $P\sim 1$~s, which
was born as a millisecond pulsar of period $P_0\ll P$, even a
$50$\% uncertainty in the value of $k_2$ would lead to an uncertainty in
the value of $\alpha$ by a factor of $\sim (P/P_0)^{1/2}\gg1$!  This highlights
the importance of obtaining quantitative solutions of oblique pulsar
magnetospheres.

\section{Relationship of Spin-down and Alignment}
\label{sec:relat-spin-down}

In this section we present a useful parametrization for torque components that unifies 
the models we have so far discussed and several others. Let us
consider the magnetospheric torque vector $\boldsymbol K$ orientation with respect to the magnetic moment
of the NS $\boldsymbol \mu$. Because of the symmetry, the alignment torque vanishes for
the aligned case, $\alpha=0^\circ$. The net torque therefore is purely
due to the spin-down torque: it is
directed along $-z$ direction and is anti-parallel to $\boldsymbol\mu$.\footnote{We note that for simplicity in this section
  we do not consider the precession torque that does not change the
  obliquity angle and pulsar rotation rate.}  
Therefore, in projections along $\boldsymbol \mu$ and
perpendicular to $\boldsymbol\mu$, the torque components become: 
\begin{subequations}
 \label{eq:K0}
\begin{align}
\alpha=0^\circ\colon K_\parallel &= -A,  \\
K_\perp &= 0,
\end{align}
\end{subequations}
 where $A>0$ is a number and $K_{\parallel}$ and $K_{\perp}$ are torque components parallel
and perpendicular to $\boldsymbol\mu$, respectively, in the
$\boldsymbol\mu{-}\boldsymbol\Omega$ plane.
Let us consider an orthogonal rotator, $\alpha=90^\circ$.  Clearly, 
due to the symmetry of the orthogonal rotator
relative to $z=0$ plane,
its alignment torque also vanishes. Therefore the net torque is again
purely due to the spin-down torque and points in the $-z$ direction.
However, this time this direction is perpendicular to $\boldsymbol\mu$,
therefore:
\begin{subequations}
\label{eq:K90}
\begin{align}
\alpha=90^\circ\colon K_\parallel &= 0,\\
K_\perp &= -B, 
\end{align}
\end{subequations}
where $B>0$ is a number.
Clearly, a simple dependence on $\alpha$ that satisfies the limiting cases
\eqref{eq:K0} and \eqref{eq:K90} is
\begin{subequations}
\label{comp}
\begin{align}
\text{Any $\alpha$:}  \quad K_{\parallel} &= - A\cos \alpha,\\
K_{\perp} &= - B \sin \alpha.
\end{align}
\end{subequations}

So far we neglected any obliquity-angle dependence of coefficients
$A$ and $B$. This is a good approximation in the simplest cases.  For
example, for vacuum pulsars the magnetospheric torque is
always perpendicular to the magnetic moment, with 
\begin{align}
  \label{eq:ABvacuum}
\text{Vacuum:} \quad A&=0, \\
B&=2 K_{\rm aligned}/3.
\end{align}
In general, however, coefficients $A$ and $B$ \emph{can} show some
obliquity angle dependence that is specific to an underlying
magnetosphere model, so from now on we will implicitly assume that 
$A=A(\alpha)$, $B=B(\alpha)$.  

A simple coordinate transformation gives us (see also Fig.~\ref{fig:axes})
\begin{align}
K_{x} &= K_{\parallel}\sin\alpha - K_{\perp}\cos\alpha\\
K_{z} &= K_{\parallel}\cos\alpha + K_{\perp}\sin\alpha.
\end{align}
Substituting for $K_{\parallel}$ and $K_{\perp}$ according to equations (\ref{comp}), we find
\begin{align}
K_{x} &= (B - A) \sin \alpha \cos \alpha,\\
K_{z} &= -A\cos^2 \alpha - B\sin^2 \alpha,
\end{align}
where, according to eqs.~\eqref{eq:K0} and \eqref{eq:K90}, $A|_{\alpha=0^{\circ}}=-K_z|_{\alpha=0^{\circ}}$ and
$B|_{\alpha=90^{\circ}}=-K_z|_{\alpha=90^{\circ}}$ give the spin-down
torque for aligned and orthogonal rotators, respectively. 
Using eqs.~\eqref{ff1}--\eqref{ff2}, for our MHD model we find that 
\begin{align}
\text{MHD:}\quad A &=K_{\rm aligned}\left[k_0+(k_1 - k_2)\sin^2\alpha\right],\\
B &= K_{\rm aligned}\left[k_0 + k_2 +(k_1 - k_2)\sin^2\alpha\right]. 
\end{align}
Note that since we approximately have $k_1 \approx k_2$, the
coefficients $A$ and $B$ for the MHD pulsar magnetosphere are
approximately independent of $\alpha$, just like in the case of the vacuum
pulsar discussed above.

Is there any relationship between the processes of alignment and spindown of the
star? If coefficients $A$ and $B$ are independent of obliquity angle $\alpha$
(this is the case for vacuum pulsars and for MHD pulsars if
$k_1\approx k_2$), the alignment and spindown torque components are related as
\begin{equation}
K_x = (B-A) \sin \alpha \cos \alpha \equiv \left[K_z(0^{\circ})-K_z(90^{\circ})\right] \sin \alpha \cos \alpha.
\end{equation}
Since ${{\rm d}\alpha}/{{\rm d} t}\propto -K_x$ (see
\S\ref{sec:torques}), and $K_z$ is a monotonic function of the obliquity
angle, this relation shows that for any $A$ and $B$ we have:
\begin{equation}
\frac{{\rm d}\alpha}{{\rm d} t}\propto \frac{{\rm d }K_z}{{\rm d}
  \alpha} \propto -\frac{{\rm d }}{{\rm d} \alpha}\left|I\frac{{\rm
      d}\Omega}{{\rm d} t}\right|,
\end{equation}
so the pulsar obliquity angle evolves in a way to reduce pulsar
spindown losses. Since in both
vacuum and MHD scenarios we have $A<B$, pulsar obliquity angle evolves
toward zero (aligned pulsar). In the special case, $A=B$, the spindown
losses are independent of the obliquity angle. In this case,
$\alpha$ does not evolve in time.  This is the case of the
force-free split-monopole solution discussed in
\citet{Bogovalov}. Finally, if we had $A>B$, the pulsar would have
evolved toward $\alpha = 90^\circ$ (orthogonal pulsar).  This result was found by \citet{BGI} and is sensitive to their \emph{assumption} about the magnitude of the magnetospheric current, a crucial yet unknown quantity at the time of their work. The advent of force-free \citep{spit06} and MHD \citep{SashaMHD} magnetospheric models for oblique pulsar magnetospheres demonstrated that the magnitude of the magnetospheric current is substantially different from that assumed by \citet{BGI} (see Appendix~\ref{app:bgi} for more detail).

\section{Discussion and Conclusions}
\label{sec:disc-concl}
It is well-understood that pulsar magnetospheres are filled
with plasma that produces the observed radio through gamma-ray emission. 
In this paper, we for the first time use self-consistent plasma-filled
solutions of pulsar magnetospheres to study the time evolution of pulsar 
obliquity angle, $\alpha$. For a spherically-symmetric NS embedded
with a magnetic dipole moment, we show that general symmetry considerations require
that pulsar obliquity angle evolves
such as to minimise pulsar spindown luminosity. Specialising to
plasma-filled (force-free or MHD) pulsars, we obtain the expressions for
magnetospheric spindown and alignment torques and provide the first self-consistent
scenario for the time evolution of the pulsar obliquity angle.
We find that
plasma-filled
pulsars approach alignment as a power-law in time, whereas vacuum
pulsars approach alignment exponentially fast.  We demonstrate that the cause for the
exponential 
decay in $\alpha$ is the zero spindown luminosity of aligned vacuum
pulsars that results in the freeze-out of the pulsar rotational period (see the
end of \S\ref{sec:ff-mhd-magn} for a discussion).

In this paper we assumed that the NS is spherically symmetric. More detailed calculations
that account for non-sphericity of the NS need to be carried
out in the case of plasma-filled magnetospheres. Such calculations are presently intrinsically uncertain due to
the lack of a detailed understanding of the NS internal structure\footnote{If the crustal contribution to the dynamical oblateness
  wins over the oblateness due to the internal magnetic field, then it
  is possible that no alignment can occur. This is more likely to be the case 
  at small pulsar periods
  \citep{1970ApJ...160L..11G} and will be the subject of a future investigation.}. In this light, it is
encouraging that the MHD evolution of pulsar obliquity angle in time
for a spherically-symmetric NS
does not obviously contradict any of the observables. In fact,  statistical
analysis of radio polarisation data suggests that 
magnetic and rotational axes align on a
time-scale of $10^6$--$10^7$~yr (\citealt{tauris98}; \citealt{young}), which is consistent with our
predictions (see Fig.~\ref{fig:evolution}).
The MHD evolution of pulsar obliquity angle also
naturally explains the production of normal pulsars (with periods of
order of one second). This is in contrast to the vacuum pulsar
scenario in which, absent stellar non-sphericity,
the production of normal pulsars is not possible due to the exponential decay
of the obliquity angle that causes the freeze-out of pulsar period
\citep{1970ApJ...160L..11G,Melatos}.
Recently, \citet{Crabpaper} proposed that time evolution of the
  radio emission profile of the Crab pulsar can be interpreted as due
  to the increase of
  the obliquity angle. This could be due to
  free precession of the NS caused by stellar non-sphericity, or it could be
  due to deviations of the stellar field from the dipolar shape. 
These possibilities will be explored in future work.

The quantitative treatment of spin-down and obliquity angle evolution
is of great importance for pulsar population synthesis studies.
It
was previously shown that the observed
fraction of nearly orthogonal rotators, which show radio interpulses,
cannot be reproduced without accounting for obliquity
angle evolution  (see, e.g., \citealt{welt}). Our plasma-filled model
may self-consistently resolve this issue.

Until now, most of the work in the area of radiopulsar population
synthesis has been performed under assumption of magnetodipole losses (see
\citealt{popov} for review, see also \citealt{kaspi} and
\citealt{ridley}). The 
self-consistent physically-motivated plasma-filled model presented in this work 
can be used to construct more realistic 
pulsar population synthesis models.

Since our computational star is much larger than the real pulsar (in
our models $0.1\le R_*/\Rlc\le0.5$ whereas for the real pulsar
$R_*/\Rlc
\approx 0.02 \times (P/10\,{\rm ms})^{-1}$), we
study for the first time how magnetospheric torques depend on the
relative size of the star
compared to the size of the light cylinder in the case of
plasma-filled magnetospheres. We find that smaller stars lose their
energy slightly more efficiently and that the spindown and 
alignment torques for $R_*/\Rlc=0.1$
are close to their asymptotic values at $R_*/\Rlc\to0$ [see
Fig.~\ref{fig:k1} and eqs.~\eqref{eq:k1}--\eqref{eq:k2}]. The origin
of this dependence on the stellar radius will
be studied elsewhere.
 
In this work we neglected stellar multipole magnetic moments higher than the dipole
moment. While this is expected to be a good approximation due to the
rapid decay of higher order magnetic moments with radius, the residual
contribution of higher-order multipoles at the distances of light cylinder can affect the
magnetospheric torques.  Importantly, since the alignment torque is
set at the surface of the star, its value might be particularly sensitive to the
higher order multipole contribution. The effect of multipolar
magnetic field configuration on pulsar evolution will be studied in
future work.

\acknowledgements

We thank J. Arons, L. Arzamasskiy, V.S. Beskin, R. Blandford,
C.-A. Faucher-Gigu\`ere, P. Goldreich,  A. Jessner, R. Narayan,
A. Spitkovsky, T. Tauris, D. Uzdensky and J. Zrake for insightful discussions. AT was supported by a Princeton Center for Theoretical Science
Fellowship and by NASA through the Einstein Fellowship Program, grant
PF3-140115. The
simulations presented in this article used computational resources
supported by the PICSciE-OIT High Performance Computing Center and
Visualisation Laboratory, and by XSEDE allocation TG-AST100040 on 
NICS Kraken and Nautilus and TACC Lonestar, Longhorn and Ranch. 

\appendix

\section{Magnitude of magnetospheric current and evolution of pulsar obliquity angle}
\label{app:bgi} 
In this paper, in order to compute the magnetospheric torques on the NS, we first calculate fluxes of various components of angular momentum. Then, by evaluating those components at the surface of the star, we compute the torque components according to eqs.~\eqref{eq:Ksph}. An alternative approach, used by, e.g., \citet{BGI}, is to calculate the torque associated with the Lorentz force
arising from the interaction of the NS poloidal field with the surface currents (these currents close the
longitudinal currents flowing in the region of open magnetosphere).  One
can prove by straightforward but cumbersome calculation that the two approaches are identical. 

As we explained in \S\ref{sec:relat-spin-down}, \citet{BGI} conclude that pulsar obliquity angle evolves in time toward an orthogonal configuration, which is the opposite of our findings (\S\ref{sec:relat-spin-down}). What is the origin of this discrepancy?
A major difference of our work and theirs
is that in the absence of a global plasma-filled magnetospheric solution at the time \citet{BGI} \emph{assumed} that the magnetospheric current
approximately equals the Goldreich-Julian (GJ) current, $j \approx
j_{\rm GJ} = \rho_{\rm GJ}c$, for all values of obliquity, where
$\rho_{\rm GJ}$ is the charge density, which is required to screen the
electric field component parallel to the magnetic field and is called
the Goldreich-Julian charge density \citep{gol69}. This is a crucial assumption because pulsar spindown luminosity is proportional to the magnetospheric current squared. Since $j_{\rm GJ}$ is
much smaller for an orthogonal rotator than for an aligned rotator,
within the assumption of \citet{BGI} the energy losses are also much smaller
for the orthogonal rotator, leading to pulsar obliquity angle evolution towards the orthogonal configuration (see \S\ref{sec:relat-spin-down}).

While  for an aligned rotator the assumption of \citet{BGI} was proven accurate,
for an orthogonal rotator an ideal plasma-filled (e.g., ideal force-free or MHD) magnetospheric structure requires a much larger current, $j\gg j_{\rm GJ}$.
Thus, the
spindown energy losses of a plasma-filled orthogonal rotator are much larger than for a plasma-filled aligned rotator. As discussed in \S\ref{sec:relat-spin-down}, the obliquity angle of a pulsar evolves so as to minimise the spindown energy losses. Therefore, in this case the pulsar evolves in time toward an aligned configuration.
Recent simulations of pair production in the inner gap \citep{Timpair} suggest that the microphysics
of the cascade near the polar cap can support the large currents ($j\gg j_{\rm GJ}$) required by the global
magnetospheric structure.

In a similar spirit, in order to take plasma effects into account,
\citet{Barsukov} \emph{assumed} the magnetospheric torque to be the
sum of vacuum torque and torque due to GJ current. This likely is a
reasonable approximation for an aligned pulsar since the vacuum torque
vanishes and the net torque is entirely due to the magnetospheric
current, which is close to the GJ value, $j\approx j_{\rm GJ}$. However,
this prescription likely breaks down for oblique pulsars, in which the
magnetospheric current significantly deviates from the GJ value,
$j\gg j_{\rm GJ}$.  In addition, the nonlinearity of the problem makes
it unlikely that the net torque is given by a simple sum of the
torques in vacuum and plasma-filled limits.

{\small

}

\label{lastpage}
\end{document}